\documentclass[reprint,
 amsmath,amssymb,
 aps,
prstab
]{revtex4-1}

\usepackage{graphicx}
\usepackage{dcolumn}
\usepackage{bm}

\begin{document}

\preprint{APS/123-QED}

\title{Residual Dispersion in a Combiner Ring}

\affiliation{The University of Lancaster and Cockcroft Institute, Lancaster LA1 4YW, United Kingdom}
\affiliation{The University of Manchester and Cockcroft Institute, Manchester M13 9PL, United Kingdom}
\affiliation{STFC Daresbury Laboratory and Cockcroft Institute, Daresbury Science and Innovation Campus, Warrington, WA4 4AD, United Kingdom}

\author{Robert Apsimon\email{robert.apsimon@lancaster.ac.uk}} \affiliation{The University of Lancaster and Cockcroft Institute, Lancaster, United Kingdom}
\author{Jakob Esberg} \affiliation{CERN, Switzerland}
\author{Hywel Owen} \affiliation{The University of Manchester and Cockcroft Institute, Manchester, United Kingdom}

\date{\today}

\begin{abstract}
In this paper we present a proof to show that there exists no system of linear or nonlinear optics which can simultaneously close multiple local orbit bumps and dispersion through a single beam transport region. The second combiner ring in the CLIC drive beam recombination system, CR2, is used as an example of where such conditions are necessary. We determine the properties of a lattice which is capable of closing the local orbit bumps and dispersion and show that all resulting solutions are either unphysical or trivial.
\end{abstract}

\maketitle

\section{\label{sec:introduction}Introduction}

Typical local orbit bumps in beam transport systems vary on the timescale of 0.1-100 s and therefore use conventional dipole magnets to vary the amplitude of the orbit bump. Faster orbit bumps can be achieved with the use of kicker magnets which may operate on timescales of 10 ns up to 100 ms. Such systems can be designed to correct the dispersion function either side of the local orbit bump with relative ease. For some applications, such as the injection into the second combiner ring CR2 for the CLIC drive beam recombination system, multiple local orbit bumps are required on sub-nanosecond timescales; thus RF deflectors are required rather than conventional dipole magnets or kicker magnets.

\begin{figure}
	\centering
		\includegraphics*[width=0.5\textwidth]{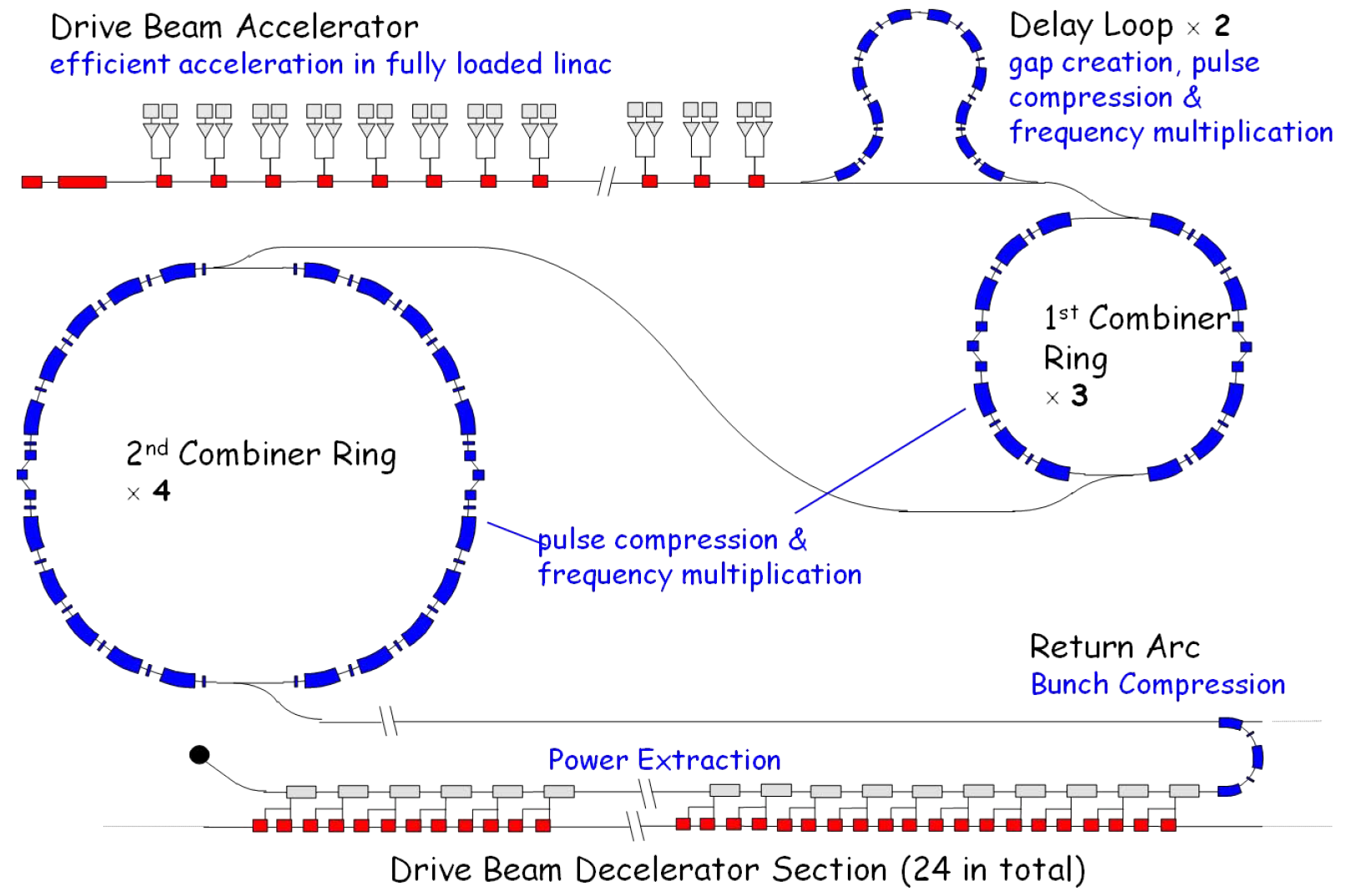}
	\caption{A schematic diagram of the CLIC drive beam recombination system \cite{CLICCDR}.}
	\label{fig:recomb}
\end{figure}

The CLIC drive beam requires 2$\times$24 pulses, each consisting of 2904 bunches with a bunch spacing of 82 ps. To achieve this, the CLIC drive beam linac produces 24$\times$24 sub-pulses with a bunch spacing of 2 ns. A recombination system is used to interleave bunches over 3 stages to produce the required pulse trains (Figure~\ref{fig:recomb}). Further details of this system can be found in \cite{CLICCDR}.

\begin{figure}
	\centering
		\includegraphics*[width=0.50\textwidth]{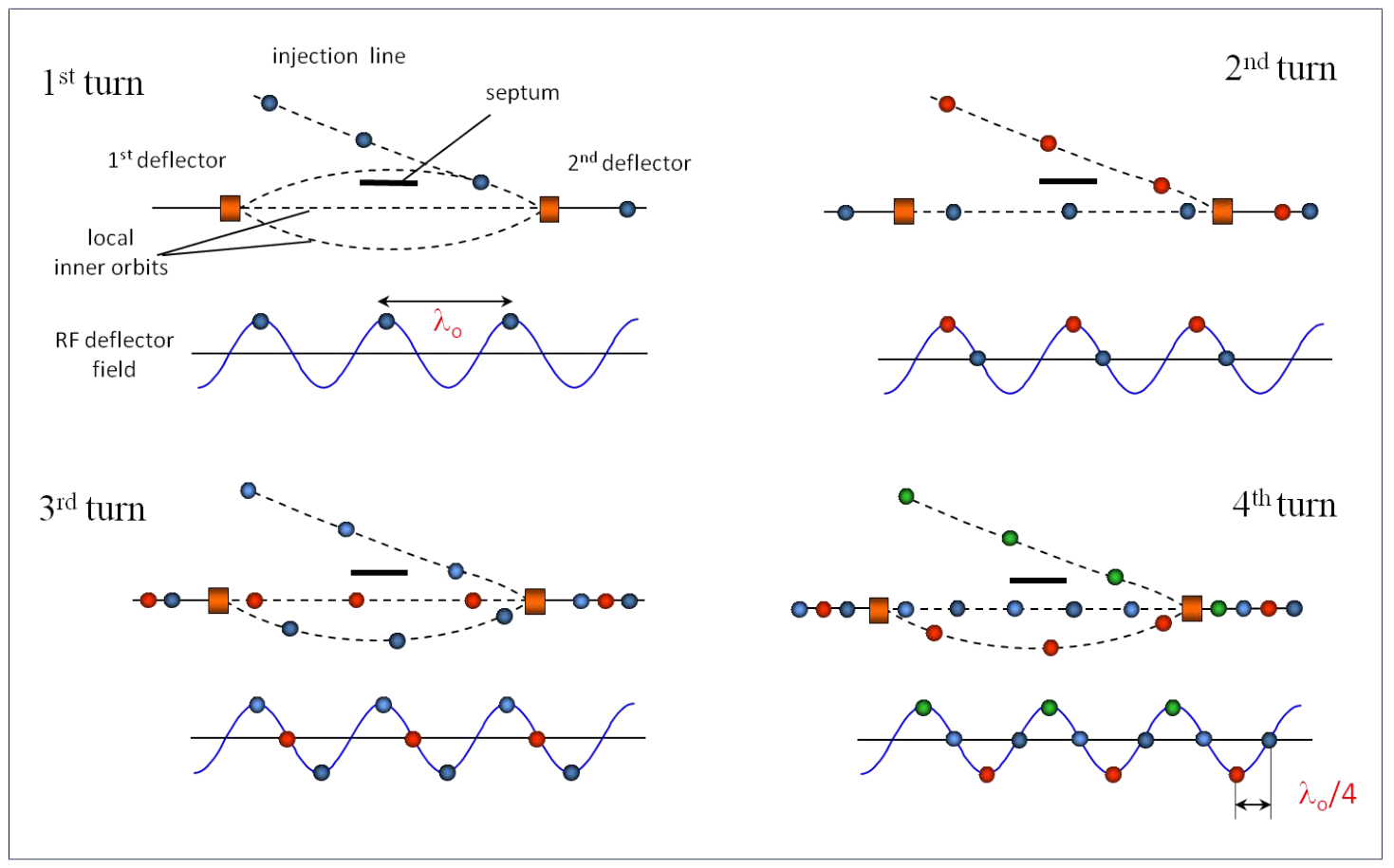}
	\caption{A schematic diagram to show how the combiner ring injection region interleaves bunches over 4 turns \cite{CLICCDR}. On each turn the stored bunches take different trajectories.}
	\label{fig:combiner}
\end{figure}

The second combiner ring stores bunch trains for up to 3.5 turns; on each turn an additional bunch train is injected such that the bunches are interleaved with the stored bunches. The principle of the injection scheme is depicted in Figure~\ref{fig:combiner}; as is shown, there are two stored trajectories and the injection trajectory passing at the same time through the injection region. In order to avoid beam losses at the injection septum magnet a bump amplitude of $\sim$3~cm is required and to interleave bunches with a bunch spacing of 82~ps (12~GHz) a 3~GHz RF deflector is required with the bunches $90^{\circ}$ apart in RF phase.

\begin{figure}
	\centering
		\includegraphics*[width=0.50\textwidth]{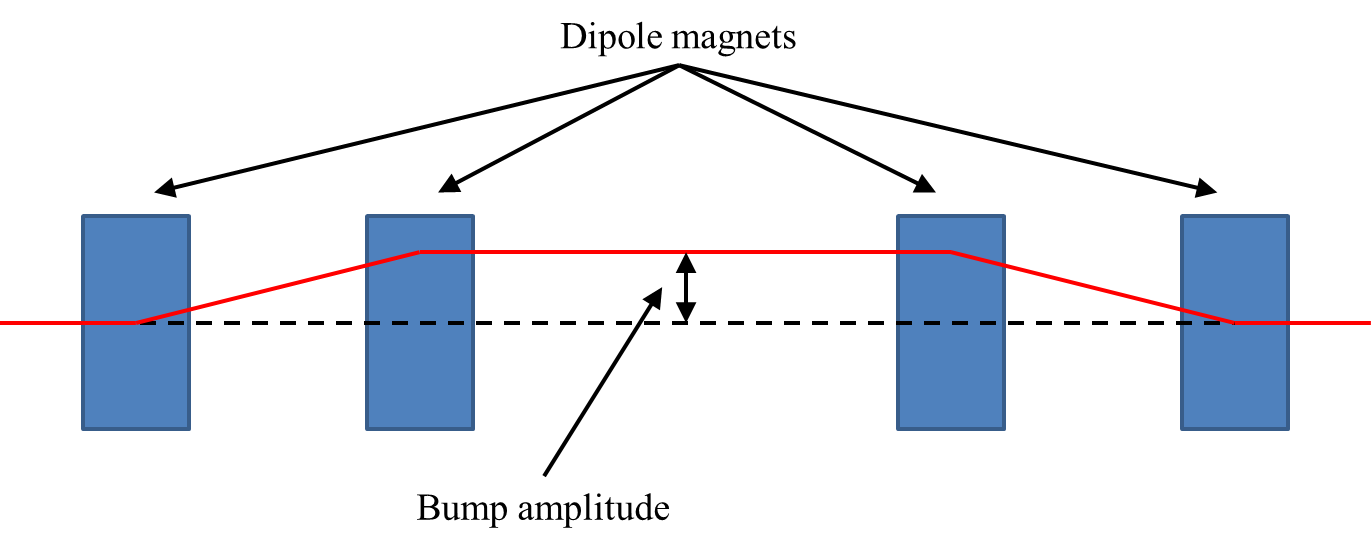}
	\caption{A schematic diagram of a 4-bump.}
	\label{fig:4bump}
\end{figure}

\begin{figure}
	\centering
		\includegraphics*[width=0.50\textwidth]{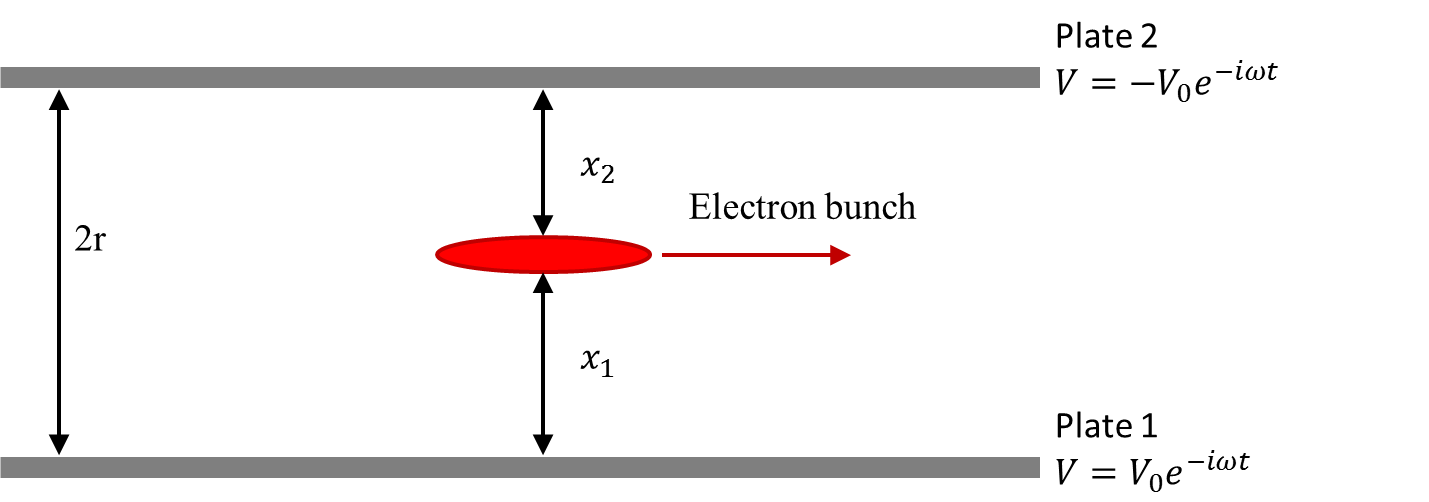}
	\caption{A schematic diagram of an RF deflector.}
	\label{fig:rfdeflector}
\end{figure}

A conventional orbit bump can be achieved with the use of 4 dipole magnets to create a dispersion-free 4-bump (Figure~\ref{fig:4bump}). However, for the CR2 injection system a 4-bump is not possible because the bump amplitude is comparable with the wavelength of the deflector RF ($\lambda = 10$~cm); this causes transverse variations in the deflecting field. To show this, we can consider the RF deflector as two parallel conducting plates separated by a distance $2r$ where the beam passes through the deflector a distance $x_{1}$ and $x_{2}$ from plates 1 and 2 respectively (Figure~\ref{fig:rfdeflector}). The phase difference of the RF on the two plates is $180^{\circ}$, thus there is a time-varying electric field across the gap.

We can assume that the deflector plates emit a time-varying electric field of the form $E_{0}e^{-i\left(\omega{t} + kx\right)}$. Hence the electric field experienced by the electron bunch is

\begin{equation}
E_{beam} = \frac{V_{0}}{x_{1}}e^{-i\left(\omega{t} + kx_{1}\right)} - \frac{V_{0}}{x_{2}}e^{-i\left(\omega{t} - kx_{2}\right)}.
\label{eq:rfdeflector}
\end{equation}

If $x_{1} = x_{2} = r$, the beam passes through the centre of the deflector and the electric field is

\begin{equation}
E_{beam} = \frac{2V_{0}}{r}\sin\left(kr\right)e^{-i\omega{t}}.
\label{eq:rfdeflector1}
\end{equation}

If $x_{1} \neq x_{2}$, the electric field experienced by the electron bunch has an apparent shift in phase and amplitude. Therefore an RF 4-bump can only be closed for one trajectory; this is not suitable for the CR2 injection region as there are two stored trajectories which need to form closed orbit bumps.

\begin{figure}
	\centering
		\includegraphics*[width=0.50\textwidth]{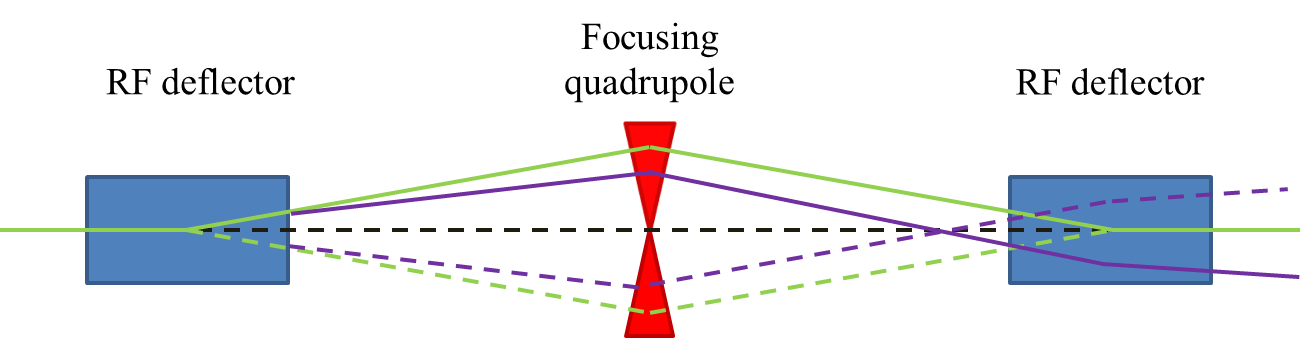}
	\caption{A schematic diagram of a local orbit bump (green) and dispersion (purple) with two RF deflectors and a single focusing quadrupole. The solid and dashed green lines show equal and opposite amplitude orbit bumps through the region.}
	\label{fig:1qbump}
\end{figure}

The orbit bumps in the injection region could be closed with two RF deflectors and a lattice of multipoles (such as quadrupoles) as depicted in Figure~\ref{fig:1qbump}. If the beam were to travel on-axis through the quadrupole, this lattice would be an achromat and would be dispersion-free; however the dipole term due to traveling off-axis through the quadrupole gives a contribution to the disperion, which prevents the dispersion closing through the lattice. As will be shown, there exists no system of linear or nonlinear optics between the RF deflectors which can simultaneously correct both the dispersion and the orbit bump.

\section{Requirements for a linear solution}

To prove that there is no linear solution for an off-axis local orbit bump which can simultaneously close both the orbit bump and the dispersion, we will first assume that there is a solution, determine the properties of such a lattice and then show that the required properties are either unphysical or trivial.

\begin{figure}
\begin{center}
\begin{tabular}{c}
\includegraphics*[width=0.50\textwidth]{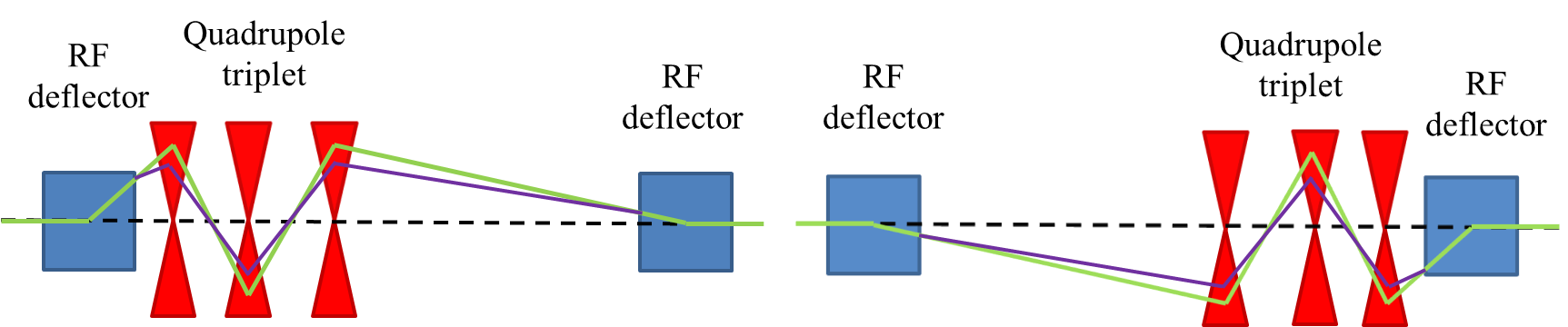}\\
{(a)}\\
\includegraphics*[width=0.50\textwidth]{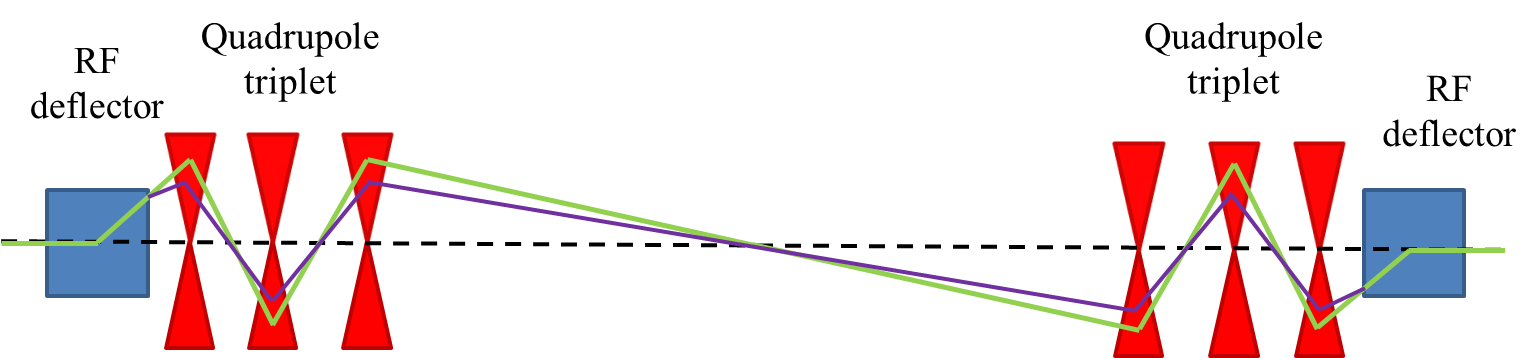}\\
{(b)}
\end{tabular}
\end{center}
\caption{Diagrams illustrating the orbit deviation (green) and dispersion (purple) to show that the reflection of an asymmetric solution is also a solution (a) and that the two can be joined to form a symmetrised solution (b).}
\label{fig:symmcell}
\end{figure}

If a linear solution exists, then it must be possible to construct a symmetric lattice which is a solution. To verify this, let us consider a hypothetical asymmetric lattice which is a solution; the reflection of this lattice must also be a solution (Figure~\ref{fig:symmcell}(a)). If we connect the original lattice to its reflection, remove the two RF deflectors in the centre and correct the central drift length, then this new lattice will also be a solution and will be symmetric (Figure~\ref{fig:symmcell}(b)). We define this as the symmetrised solution \cite{CLICnote}.

To verify that an asymmetric lattice can be symmeterised, we will consider the orbit deviation and dispersion introduced by a magnetic dipolar kick; these can be expressed respectively as shown in Eq.~\ref{eq:symmintro1} where $\rho$ is the radius of curvature and $\theta$ is the deflection angle.

\begin{equation}
\begin{matrix}
\begin{pmatrix} x \\
x' \end{pmatrix}=
\begin{pmatrix} \rho\left(1-\cos\theta\right) \\
\tan\theta \end{pmatrix} \\ \\
\begin{pmatrix} D_{x} \\
D'_{x} \end{pmatrix}=
\begin{pmatrix} \rho\left(1-\cos\theta\right) \\
\sin\theta \end{pmatrix}
\end{matrix}
\label{eq:symmintro1}
\end{equation}

From Eq.~\ref{eq:symmintro1}, we can express the dispersion in terms of the orbit deviation

\begin{equation}
\begin{pmatrix} D_{x} \\
D'_{x} \end{pmatrix}=
\begin{pmatrix} x \\
x'\cos\theta \end{pmatrix}.
\label{eq:symmintro2}
\end{equation}

It should be noted that this relationship holds true for all electromagnetic dipolar kicks. In order to close the orbit and dispersion through a dipole, the incident trajectory and dispersion must be

\begin{equation}
\begin{matrix}
\begin{pmatrix} x_{0} \\
x'_{0} \end{pmatrix}=
\begin{pmatrix} x \\
-x' \end{pmatrix} \\ \\
\begin{pmatrix} D_{x,0} \\
D'_{x,0} \end{pmatrix}=
\begin{pmatrix} D_{x} \\
-D'_{x} \end{pmatrix}=
\begin{pmatrix} x \\
-x'\cos\theta \end{pmatrix}
\end{matrix}.
\label{eq:symmintro3}
\end{equation}

If we imagine removing the dipole and replacing it with a drift length, we can calculate the required drift length to obtain $x=0$ and $D_{x}=0$ respectively

\begin{equation}
\begin{matrix}
L_{x=0} = -\frac{x_{0}}{x'_{0}} = \frac{x}{x'} \\
L_{D_{x}=0} = -\frac{D_{x,0}}{D'_{x,0}} = \frac{x}{x'\cos\theta}=\frac{L_{x=0}}{\cos\theta}
\end{matrix}.
\label{eq:symmintro4}
\end{equation}

Therefore the orbit deviation and dispersion equal zero at different locations. However if we consider the midpoint of the symmeterised lattice to be where $D_{x}=0$ and calculate the orbit deviation and dispersion a distance $L_{D_{x}=0}$ downstream of the midpoint, we obtain

\begin{equation}
\begin{matrix}
\begin{pmatrix} x_{1} \\
x'_{1} \end{pmatrix}=
\begin{pmatrix} x_{0} - \frac{2x_{0}}{\cos\theta} \\
x'_{0} \end{pmatrix}=
\begin{pmatrix} -x_{0} - \frac{2x_{0}\left(1-\cos\theta\right)}{\cos\theta} \\
x'_{0} \end{pmatrix} \\ \\
\begin{pmatrix} D_{x,1} \\
D'_{x,1} \end{pmatrix}=
\begin{pmatrix} -D_{x,0} \\
D'_{x,0} \end{pmatrix}
\end{matrix}.
\label{eq:symmintro5}
\end{equation}

\begin{figure}
	\centering
		\includegraphics*[width=0.50\textwidth]{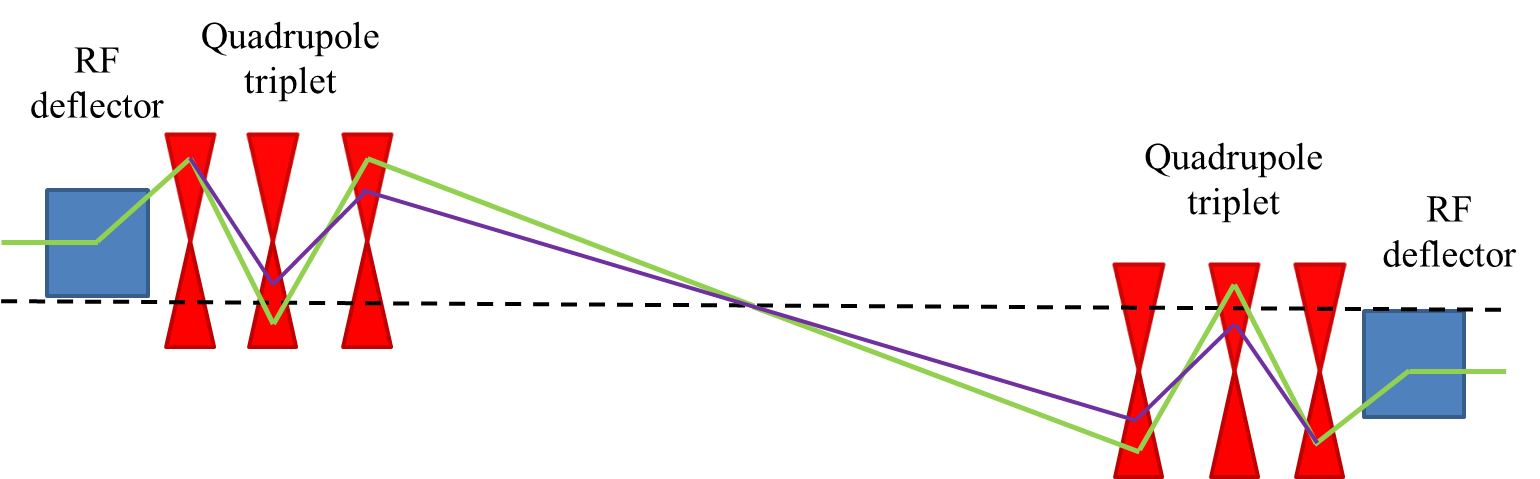}
	\caption{A schematic diagram of the orbit deviation (green) and dispersion (purple) through a symmeterised lattice depicting the transverse offset in quadrupoles required to symmeterise an asymmetric lattice.}
	\label{fig:offset}
\end{figure}

As will be shown later, the dispersion function fulfills the condition for an anti-symmetric solution. If we apply a transverse offset of $-\frac{2x_{0}\left(1-\cos\theta\right)}{\cos\theta}$ to the multipoles in the second half of the symmeterised lattice with respect to the first half of the cell (Figure~\ref{fig:offset}), then the orbit deviation also becomes an anti-symmetric solution.

\begin{figure}
	\centering
		\includegraphics*[width=0.50\textwidth]{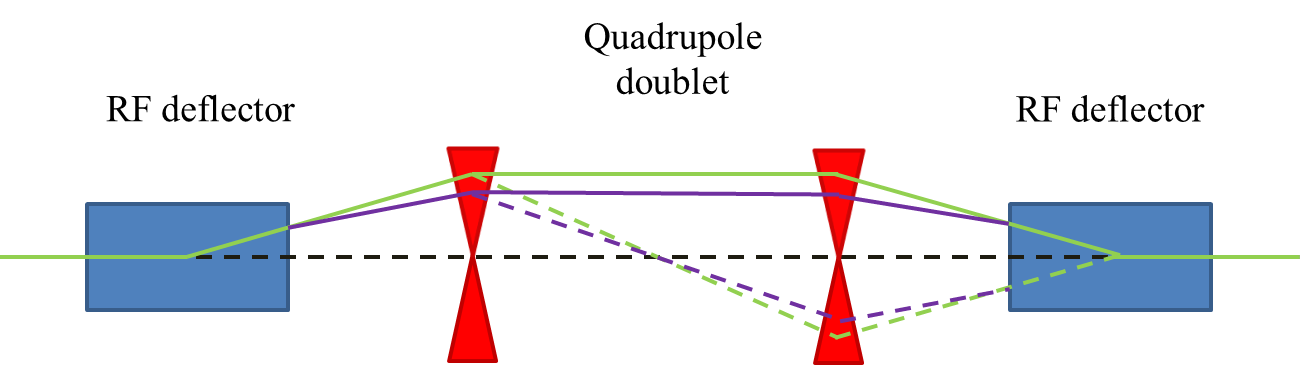}
	\caption{A schematic diagram of the orbit deviation (green) and dispersion (purple) depicting the symmetric (solid lines) and anti-symmetric (dashed lines) solutions through a quadrupole doublet.}
	\label{fig:parity}
\end{figure}

Having shown that a symmetric lattice must exist if any solution exists, we are able to greatly simplify the problem. As the lattice is symmetric about the midpoint and closes the orbit bump and dispersion, this implies that the orbit bump and dispersion function must also be symmetric or anti-symmetric (Figure~\ref{fig:parity}); we will define these as the even and odd parity solutions respectively. It should be noted that the orbit and dispersion must have the same parity because the dispersion through a multipole is dependent on its trajectory; with the exception of dipoles. However dipoles break the transverse symmetry of a lattice; thus it is not possible to close more than one orbit bump.

\begin{figure}
\begin{center}
\begin{tabular}{c}
\includegraphics*[width=0.50\textwidth]{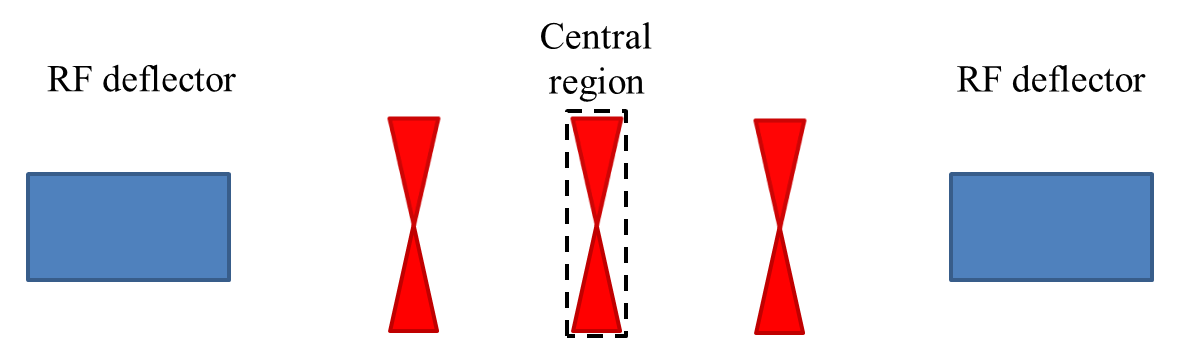}\\
{(a)}\\
\includegraphics*[width=0.50\textwidth]{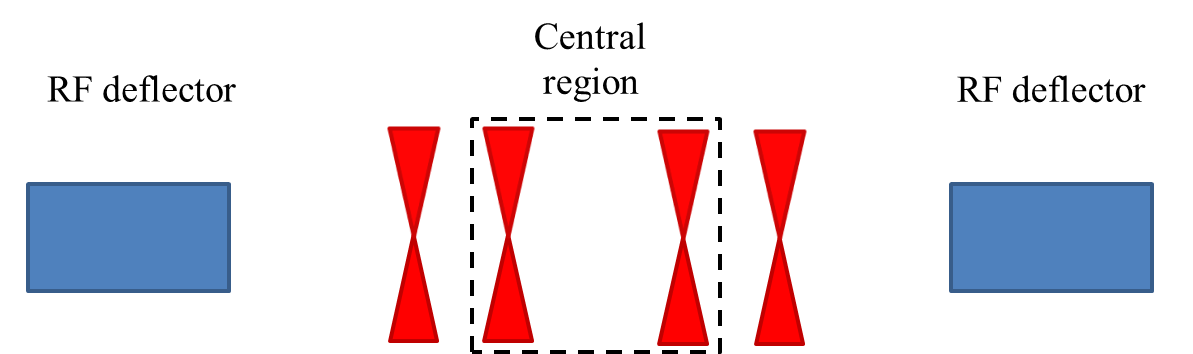}\\
{(b)}
\end{tabular}
\end{center}
\caption{Diagrams to show the central region for a symmetric lattice with an odd number of multipoles (a) and an even number of multipoles (b).}
\label{fig:centralregion}
\end{figure}

For a symmetric lattice, we define the `central region' as the central multipole if there are an odd number of multipoles in the lattice and the central doublet if there are an even number of multipoles (Figure~\ref{fig:centralregion}).

As previously stated, if a symmetric lattice closes both the local orbit bump and dispersion, the orbit and dispersion must also be symmetric with odd or even parity. Thus the bump trajectory and dispersion must also be symmetric across the central region. Therefore we need only investigate the central region of a hypothetical lattice to make general conclusions about the criteria for a solution for all possible lattice designs. If we can verify that no solution exists for the singlet and doublet cases then by induction, we can conclude that there is no solution for any symmetric sequence of $2n+1$ quadrupoles and $2n$ quadrupoles respectively; hence no symmetric sequence of quadrupoles can be a solution. If no symmetric solution exists then from our earlier statement, no linear solution exists.

The singlet can be considered as a special case of a doublet, where the drift length between the quadrupoles is zero. Thus we need only consider the doublet case to determine if a solution exists for either case.

\section{Beam dynamics}

We recall that the transfer matrices for focussing and defocussing quadrupoles are

\begin{equation}
\begin{matrix}
\mathbf{M}_{qf}=\begin{pmatrix} \cos\left(\sqrt{k_{f}}l_{q}\right) & \frac{\sin\left(\sqrt{k_{f}}l_{q}\right)}{\sqrt{k_{f}}} \\
-\sqrt{k_{f}}\sin\left(\sqrt{k_{f}}l_{q}\right) & \cos\left(\sqrt{k_{f}}l_{q}\right) \end{pmatrix} \\ \\
\mathbf{M}_{qd}=\begin{pmatrix} \cosh\left(\sqrt{k_{d}}l_{q}\right) & \frac{\sinh\left(\sqrt{k_{d}}l_{q}\right)}{\sqrt{k_{d}}} \\
\sqrt{k_{d}}\sinh\left(\sqrt{k_{d}}l_{q}\right) & \cosh\left(\sqrt{k_{d}}l_{q}\right) \end{pmatrix}\end{matrix}
\label{eq:quad1}
\end{equation}

\noindent and for a drift space that the matrix is

\begin{equation}
\mathbf{M}_{dr}=\begin{pmatrix} 1 & L_{dr} \\
0 & 1 \end{pmatrix}.
\label{eq:quad2}
\end{equation}

The trajectory downstream of the central region can be defined in terms of the trajectory upstream

\begin{equation}
\begin{pmatrix} x_{1} \\
x'_{1} \end{pmatrix}=\mathbf{M}
\begin{pmatrix} x_{0} \\
x'_{0} \end{pmatrix},
\label{eq:quad3}
\end{equation}

\noindent where $\mathbf{M}$ is the transfer matrix for the central region. It should be noted that for any linear transfer matrix $\mathbf{M}$ describing a system where energy and momentum are conserved, $\det\left(\mathbf{M}\right)=1$.

As the beam is traveling off-axis through the quadrupoles in the central region, we need to consider the quadrupole contributions to the dispersion function. The dispersion, $D_{q}$, and its derivative, $D'_{q}$, can be defined as

\begin{equation}
\begin{aligned}
D_{q}=\mathbf{M}_{1,2}\int^{l_{q}}_{0}{\frac{\tilde{\mathbf{M}}_{1,1}}{\rho\left(s\right)}ds}-\mathbf{M}_{1,1}\int^{l_{q}}_{0}{\frac{\tilde{\mathbf{M}}_{1,2}}{\rho\left(s\right)}ds} \\
D'_{q}=\mathbf{M}_{2,2}\int^{l_{q}}_{0}{\frac{\tilde{\mathbf{M}}_{1,1}}{\rho\left(s\right)}ds}-\mathbf{M}_{2,1}\int^{l_{q}}_{0}{\frac{\tilde{\mathbf{M}}_{1,2}}{\rho\left(s\right)}ds}
\end{aligned},
\label{eq:quad4}
\end{equation}

\noindent where $\tilde{\mathbf{M}}_{i,j}=\mathbf{M}_{i,j}\left(s\right)$ and $\rho\left(s\right)$ is the radius of curvature at a longitudinal position $s$ in the quadrupole, given as

\begin{equation}
\rho\left(s\right)=\frac{dL}{d\theta}=\frac{\left(1+x'^{2}\right)^{\frac{3}{2}}}{x''}.
\label{eq:quad5}
\end{equation}

By substituting the results of Eq. \ref{eq:quad3} and Hill's Equation, $x''+kx=0$, into Eq. \ref{eq:quad5}, the radius of curvature can be obtained as

\begin{equation}
\rho\left(s\right)=\frac{\left(1+\left(\tilde{\mathbf{M}}_{2,1}x_{0}+\tilde{\mathbf{M}}_{2,2}x'_{0}\right)^{2}\right)^{\frac{3}{2}}}{x_{0}\frac{d\tilde{\mathbf{M}}_{2,1}}{ds}+x'_{0}\frac{d\tilde{\mathbf{M}}_{2,2}}{ds}}.
\label{eq:quad7}
\end{equation}

For the focusing and defocusing quadrupoles the radius of curvature is given respectively as

\begin{equation}
\begin{aligned}
\rho_{f}\left(s\right)=-\frac{\left(1+\left(\tilde{\mathbf{M}}_{2,1}x_{0}+\tilde{\mathbf{M}}_{2,2}x'_{0}\right)^{2}\right)^{\frac{3}{2}}}{k_{f}\left(\tilde{\mathbf{M}}_{1,1}x_{0}+\tilde{\mathbf{M}}_{1,2}x'_{0}\right)} \\
\rho_{d}\left(s\right)=\frac{\left(1+\left(\tilde{\mathbf{M}}_{2,1}x_{0}+\tilde{\mathbf{M}}_{2,2}x'_{0}\right)^{2}\right)^{\frac{3}{2}}}{k_{d}\left(\tilde{\mathbf{M}}_{1,1}x_{0}+\tilde{\mathbf{M}}_{1,2}x'_{0}\right)}
\end{aligned}.
\label{eq:quad8}
\end{equation}

\section{Central region}

In total there are only 8 possible scenarios for the central region which could produce a solution:

\begin{itemize}
	\item Focussing singlet, symmetric bump
	\item Focussing singlet, anti-symmetric bump
	\item Defocussing singlet, symmetric bump
	\item Defocussing singlet, anti-symmetric bump
	\item Focussing doublet, symmetric bump
	\item Focussing doublet, anti-symmetric bump
	\item Defocussing doublet, symmetric bump
	\item Defocussing doublet, anti-symmetric bump
\end{itemize}

As previously stated, the singlet can be treated as a special case of a doublet, thus allowing us to study all scenarios at once.

\subsection{Quadrupole doublet}

Rather than using specific matrix elements for a focussing or defocussing quadrupole from Eq. \ref{eq:quad1}, we will write the equations in a more general form.

For the symmetric solution, we require

\begin{equation}
\begin{matrix}
\begin{pmatrix} x_{1} \\
x'_{1} \end{pmatrix}=
\begin{pmatrix} x_{0} \\
-x'_{0} \end{pmatrix} \\ \\
\begin{pmatrix} D_{x,1} \\
D'_{x,1} \end{pmatrix}=
\begin{pmatrix} D_{x,0} \\
-D'_{x,0} \end{pmatrix}
\end{matrix},
\label{eq:bdoub1a}
\end{equation}

\noindent where point `0' refers to the start of the singlet or doublet in the central region and point `1' refers to the end. For the anti-symmetric solution we require

\begin{equation}
\begin{matrix}
\begin{pmatrix} x_{1} \\
x'_{1} \end{pmatrix}=
\begin{pmatrix} -x_{0} \\
x'_{0} \end{pmatrix} \\ \\
\begin{pmatrix} D_{x,1} \\
D'_{x,1} \end{pmatrix}=
\begin{pmatrix} -D_{x,0} \\
D'_{x,0} \end{pmatrix}
\end{matrix}.
\label{eq:bsing1b}
\end{equation}

Therefore, to generalise the constraints for both parities, we require

\begin{equation}
\begin{matrix}
\begin{pmatrix} x_{1} \\
x'_{1} \end{pmatrix}=
\begin{pmatrix} \pm x_{0} \\
\mp x'_{0} \end{pmatrix} \\ \\
\begin{pmatrix} D_{x,1} \\
D'_{x,1} \end{pmatrix}=
\begin{pmatrix} \pm D_{x,0} \\
\mp D'_{x,0} \end{pmatrix}
\end{matrix},
\label{eq:bdoub1c}
\end{equation}

\noindent where the top sign in $\pm$ or $\mp$ represents the sign for the symmetric case and the bottom sign for the anti-symmetric case. If we consider the general transfer matrix for a quadrupole to be

\begin{equation}
\mathbf{M}=\begin{pmatrix}
\mathbf{M}_{1,1} & \mathbf{M}_{1,2} \\
\mathbf{M}_{2,1} & \mathbf{M}_{2,2} \end{pmatrix}
\label{eq:bdoub2}
\end{equation}

\noindent and the transfer matrix of a doublet to be

\begin{equation}
\begin{array}{l}
\mathbf{N}=\begin{pmatrix}
\mathbf{M}_{1,1} & \mathbf{M}_{1,2} \\
\mathbf{M}_{2,1} & \mathbf{M}_{2,2} \end{pmatrix}\begin{pmatrix}
1 & L_{dr} \\
0 & 1 \end{pmatrix}\begin{pmatrix}
\mathbf{M}_{1,1} & \mathbf{M}_{1,2} \\
\mathbf{M}_{2,1} & \mathbf{M}_{2,2} \end{pmatrix}= \\ \left(\begin{smallmatrix}
\mathbf{M}_{1,1}^{2}+L_{dr}\mathbf{M}_{1,1}\mathbf{M}_{2,1}+\mathbf{M}_{1,2}\mathbf{M}_{2,1} & L_{dr}\mathbf{M}_{1,1}^{2}+2\mathbf{M}_{1,1}\mathbf{M}_{1,2} \\
L_{dr}\mathbf{M}_{2,1}^{2}+2\mathbf{M}_{1,1}\mathbf{M}_{2,1} & \mathbf{M}_{1,1}^{2}+L_{dr}\mathbf{M}_{1,1}\mathbf{M}_{2,1}+\mathbf{M}_{1,2}\mathbf{M}_{2,1} \end{smallmatrix}\right)
\end{array}.
\label{eq:bdoub3}
\end{equation}

Given the dispersive contributions, $D_{q}$ and $D'_{q}$, for a quadrupole from Eq. \ref{eq:quad4}, we can define the contributions for a doublet, $D_{q,doub}$ and $D'_{q,doub}$, as

\begin{equation}
\begin{matrix}
D_{q,doub}=\mathbf{M}_{1,1}D_{q}+\left(L_{dr}\mathbf{M}_{1,1}+\mathbf{M}_{1,2}\right)D'_{q} \\ \\
D'_{q,doub}=\mathbf{M}_{2,1}D_{q}+\left(L_{dr}\mathbf{M}_{2,1}+\mathbf{M}_{2,2}\right)D'_{q}
\end{matrix}.
\label{eq:bdoub4}
\end{equation}

This can be simplified using the results from Eq. \ref{eq:bdoub3} to produce

\begin{equation}
\begin{matrix}
D_{q,doub}=\mathbf{N}_{1,2}\int{\frac{\tilde{\mathbf{M}}_{1,1}}{\rho\left(s\right)}ds}-\mathbf{N}_{1,1}\int{\frac{\tilde{\mathbf{M}}_{1,2}}{\rho\left(s\right)}ds} \\ \\
D'_{q,doub}=\mathbf{N}_{2,2}\int{\frac{\tilde{\mathbf{M}}_{1,1}}{\rho\left(s\right)}ds}-\mathbf{N}_{2,1}\int{\frac{\tilde{\mathbf{M}}_{1,2}}{\rho\left(s\right)}ds}
\end{matrix}.
\label{eq:bdoub5}
\end{equation}

For the orbit bump, we require

\begin{equation}
\begin{matrix}
\mathbf{N}_{1,1}x_{0}+\mathbf{N}_{1,2}x'_{0}=\pm x_{0} \\ \\
\mathbf{N}_{2,1}x_{0}+\mathbf{N}_{2,2}x'_{0}=\mp x'_{0}
\end{matrix}
\label{eq:bdoub6}
\end{equation}

\noindent and the solution to these simultaneous equations is

\begin{equation}
\frac{x_{0}}{x'_{0}}=\frac{\mathbf{N}_{1,2}}{\pm1-\mathbf{N}_{1,1}}=\frac{\mp1-\mathbf{N}_{2,2}}{\mathbf{N}_{2,1}}.
\label{eq:bdoub7}
\end{equation} 

For the dispersion function we require

\begin{equation}
\begin{matrix}
\mathbf{N}_{1,1}D_{x,0}+\mathbf{N}_{1,2}D'_{x,0}+D_{q,doub}=\pm D_{x,0} \\ \\
\mathbf{N}_{2,1}D_{x,0}+\mathbf{N}_{2,2}D'_{x,0}+D'_{q,doub}=\mp D'_{x,0}
\end{matrix},
\label{eq:bdoub9}
\end{equation}

\noindent where $D_{q,doub}$ and $D'_{q,doub}$ are the dispersive contributions from the quadrupole doublet as given in Eq. \ref{eq:bdoub5}. Eq. \ref{eq:bdoub9} can be written out more explicitly as

\begin{equation}
\begin{array}{l}
\pm D_{x,0}=\mathbf{N}_{1,1}D_{x,0}+\mathbf{N}_{1,2}D'_{x,0}+\mathbf{N}_{1,2}\int{\frac{\tilde{\mathbf{M}}_{1,1}}{\rho\left(s\right)}ds} \\
-\mathbf{N}_{1,1}\int{\frac{\tilde{\mathbf{M}}_{1,2}}{\rho\left(s\right)}ds} \\ \\
\mp D'_{x,0}=\mathbf{N}_{2,1}D_{x,0}+\mathbf{N}_{2,2}D'_{x,0}+\mathbf{N}_{2,2}\int{\frac{\tilde{\mathbf{M}}_{1,1}}{\rho\left(s\right)}ds} \\
-\mathbf{N}_{2,1}\int{\frac{\tilde{\mathbf{M}}_{1,2}}{\rho\left(s\right)}ds}
\end{array}
\label{eq:bdoub10}
\end{equation}

\noindent and these equations can be rearranged and the results from Eq. \ref{eq:bdoub7} substituted to give the following equations

\begin{equation}
\begin{array}{l}
D_{x,0}=\frac{x_{0}}{x'_{0}}D'_{x,0}+\frac{x_{0}}{x'_{0}}\int{\frac{\tilde{\mathbf{M}}_{1,1}}{\rho\left(s\right)}ds}-\frac{\mathbf{N}_{1,1}}{\pm1-\mathbf{N}_{1,1}}\int{\frac{\tilde{\mathbf{M}}_{1,2}}{\rho\left(s\right)}ds} \\ \\
D_{x,0}=\frac{x_{0}}{x'_{0}}D'_{x,0}+\left(\frac{x_{0}}{x'_{0}}\pm\frac{1}{\mathbf{N}_{2,1}}\right)\int{\frac{\tilde{\mathbf{M}}_{1,1}}{\rho\left(s\right)}ds}+\int{\frac{\tilde{\mathbf{M}}_{1,2}}{\rho\left(s\right)}ds}
\end{array}.
\label{eq:bdoub11}
\end{equation}

By solving these simultaneous equations we obtain

\begin{equation}
\left(\mp1+\mathbf{N}_{1,1}\right)\int{\frac{\tilde{\mathbf{M}}_{1,1}}{\rho\left(s\right)}ds}-\mathbf{N}_{2,1}\int{\frac{\tilde{\mathbf{M}}_{1,2}}{\rho\left(s\right)}ds}=0.
\label{eq:bdoub12}
\end{equation}

By expressing the matrix elements of $\mathbf{N}$ in terms of $\mathbf{M}$ and using the fact that $\det\left(\mathbf{M}\right)=1$, Eq. \ref{eq:bdoub12} can be simplified for the symmetric and anti-symmetric cases respectively

\begin{equation}
\begin{array}{l}
\mathbf{M}_{2,1}\left(L_{dr}\mathbf{M}_{1,1}+2\mathbf{M}_{1,2}\right)\int{\frac{\tilde{\mathbf{M}}_{1,1}}{\rho\left(s\right)}ds}\\
-\mathbf{M}_{2,1}\left(L_{dr}\mathbf{M}_{2,1}+2\mathbf{M}_{1,1}\right)\int{\frac{\tilde{\mathbf{M}}_{1,2}}{\rho\left(s\right)}ds}=0 \\ \\
\left(2\mathbf{M}_{1,1}+L_{dr}\mathbf{M}_{2,1}\right)\times \\\left(\mathbf{M}_{1,1}\int{\frac{\tilde{\mathbf{M}}_{1,1}}{\rho\left(s\right)}ds}-\mathbf{M}_{2,1}\int{\frac{\tilde{\mathbf{M}}_{1,2}}{\rho\left(s\right)}ds}\right)=0
\end{array}.
\label{eq:bdoub13}
\end{equation}

For the symmetric case in Eq. \ref{eq:bdoub13}, we can express it as

\begin{equation}
\mathbf{M}_{2,1}\left(2D_{q}+L_{dr}D'_{q}\right)=0,
\label{eq:bdoub14}
\end{equation}

\noindent where $D_{q}$ and $D'_{q}$ are the dispersive contributions for one of the quadrupoles in the doublet, or half of a quadrupole for the singlet. Therefore, either $\mathbf{M}_{2,1}=0$ or $2D_{q}+L_{dr}D'_{q}=0$; the latter implies that $D'_{q}\propto D_{q}$. We can consider $D_{q}$ and $D'_{q}$ in Eq. \ref{eq:quad4} as vectors in a coordinate system where $\int^{l_{q}}_{0}{\frac{\tilde{\mathbf{M}}_{1,1}}{\rho\left(s\right)}ds}$ and $\int^{l_{q}}_{0}{\frac{\tilde{\mathbf{M}}_{1,2}}{\rho\left(s\right)}ds}$ are the coordinate bases. If $D'_{q}\propto D_{q}$, then the corresponding vectors in our abstract coordinate system must be parallel; therefore $\frac{\mathbf{M}_{1,2}}{\mathbf{M}_{1,1}}=\frac{\mathbf{M}_{2,2}}{\mathbf{M}_{2,1}}$. This implies that $\det\left(\mathbf{M}\right)=0$, but this is a contradiction because we know that $\det\left(\mathbf{M}\right)=1$. Therefore $2D_{q}+L_{dr}D'_{q}=0$ is not possible except for the trivial case and $\mathbf{M}_{2,1}=0$ is the only solution for the symmetric case.

\noindent For the focusing quadrupoles, $\mathbf{M}_{2,1}=0$ has solutions at $\sqrt{k_{f}}l_{q}=m\pi$. For the defocusing quadrupoles, $\mathbf{M}_{2,1}=0$ only has the solution $\sqrt{k_{d}}l_{q}=0$, which is trivial and can be neglected.

\noindent For the anti-symmetric case,  Eq. \ref{eq:bdoub13} can be simplified with the use of Eq. \ref{eq:quad4} and the fact that $\mathbf{M}_{1,1}=\mathbf{M}_{2,2}$

\begin{equation}
\left(2\mathbf{M}_{1,1}+L_{dr}\mathbf{M}_{2,1}\right)D'_{q}=0.
\label{eq:bdoub15}
\end{equation}

Therefore either $D'_{q}=0$ or $2\mathbf{M}_{1,1}+L_{dr}\mathbf{M}_{2,1}=0$.

For the defocusing singlet, there are no real solutions to $2\mathbf{M}_{1,1}+L_{dr}\mathbf{M}_{2,1}=0$ and for the doublet there are no solutions for $l_{q},L_{dr}>0$. For the other solution, $D'_{q}=0$, the only solution is the trivial case when $\sqrt{k_{d}}l_{q}=0$.

For the focusing singlet, solutions for $2\mathbf{M}_{1,1}+L_{dr}\mathbf{M}_{2,1}=0$ occur at $\sqrt{k_{f}}l_{q}=\left(2m+1\right)\pi/2$ and $D'_{q}=0$ when $\sqrt{k_{f}}l_{q}=m\pi$; these solution sets can be combined to give $\sqrt{k_{f}}l_{q}=m\pi/2$. For the doublet both $D'_{q}=0$ and $2\mathbf{M}_{1,1}+L_{dr}\mathbf{M}_{2,1}=0$ have non-trivial solutions only if $\tan\left(\sqrt{k_{f}}l_{q}\right)=\sqrt{k_{f}}L_{dr}/2$.

\section{Results}

From the results above, we can conclude that there are no non-trivial solutions for any of the defocusing quadrupole cases. Thus we can investigate the possible solutions for the focusing quadrupoles. There are three distinct possible solutions for the focusing quadrupole cases, which are

\begin{itemize}
	\item $\sqrt{k_{f}}l_{q}=m\pi$ for the symmetric focusing cases
	\item $\sqrt{k_{f}}l_{q}=m\pi/2$ for the anti-symmetric focusing singlet
	\item $\tan\left(\sqrt{k_{f}}l_{q}\right)=\sqrt{k_{f}}L_{dr}/2$ for the anti-symmetric focusing doublet
\end{itemize}

\subsection{$\mathbf{\sqrt{k_{f}}l_{q}=m\pi}$ for the symmetric focusing cases}

If $\sqrt{k_{f}}l_{q}=m\pi$, then the transfer matrix $\mathbf{N}$ in Eq. \ref{eq:bdoub3} can be expressed as

\begin{equation}
\mathbf{N}=\begin{pmatrix} 1 & L_{dr} \\
0 & 1 \end{pmatrix},
\label{eq:res1}
\end{equation}

\noindent thus the requirements to produce a symmetric dispersion function become

\begin{equation}
\begin{array}{l}
D_{x,0}+L_{dr}D'_{x,0}+D_{q,doub}=D_{x,0} \\
D'_{x,0}+D'_{q,doub}=-D'_{x,0}
\end{array}.
\label{eq:res2}
\end{equation}

Therefore we obtain $\frac{D_{q,doub}}{L_{dr}}=\frac{D'_{q,doub}}{2}$, which implies that $\det\left(\mathbf{N}\right)=0$ which is contradictory as we know that $\det\left(\mathbf{N}\right)=1$. Hence there are no non-trivial solutions for the focusing cases for $\sqrt{k_{f}}l_{q}=m\pi$.

\subsection{$\mathbf{\sqrt{k_{f}}l_{q}=m\pi/2}$ for the anti-symmetric focusing singlet}

For the focussing singlet, we obtain the transfer matrix

\begin{equation}
\mathbf{N}=\begin{pmatrix} \cos\left(2\sqrt{k_{f}}l_{q}\right) & \frac{\sin\left(2\sqrt{k_{f}}l_{q}\right)}{\sqrt{k_{f}}} \\
-\sqrt{k_{f}}\sin\left(2\sqrt{k_{f}}l_{q}\right) & \cos\left(2\sqrt{k_{f}}l_{q}\right) \end{pmatrix}.
\label{eq:res3}
\end{equation}

If $\sqrt{k_{f}}l_{q}=\left(2m+1\right)\pi/2$, then Eq. \ref{eq:res3} becomes $-\mathbf{I}$, where $\mathbf{I}$ is the identity matrix. By considering the resulting orbit and dispersion functions, we obtain

\begin{equation}
\begin{array}{l}
-x_{0}=-x_{0} \\
-x'_{0}=x'_{0} \\ \\
-D_{x,0}+D_{q}=-D_{x,0} \\
-D'_{x,0}+D'_{q}=D'_{x,0}
\end{array},
\label{eq:spec5}
\end{equation}

\noindent which gives the result $D_{q}=x'_{0}=0$. Similarly, if $\sqrt{k_{f}}l_{q}=m\pi$, we obtain the result $D'_{q}=x_{0}=0$.

If $D_{q}=0$ then from Eq. \ref{eq:quad4}, it can be shown that

\begin{equation}
\mathbf{M}_{1,2}\int^{l_{q}}_{0}{\frac{\tilde{\mathbf{M}}_{1,1}}{\rho\left(s\right)}ds}=\mathbf{M}_{1,1}\int^{l_{q}}_{0}{\frac{\tilde{\mathbf{M}}_{1,2}}{\rho\left(s\right)}ds}.
\label{eq:spec6}
\end{equation}

As $\sqrt{k_{f}}l_{q}=\left(2m+1\right)\pi/2$, $\mathbf{M}_{1,1}=0$ and $\mathbf{M}_{1,2}=-1$, which implies that $\int^{l_{q}}_{0}{\frac{\tilde{\mathbf{M}}_{1,1}}{\rho\left(s\right)}ds}=0$. However, if we substitute the results from Eq. \ref{eq:quad8} then we obtain the integral

\begin{equation}
\begin{array}{l}
\int^{l_{q}}_{0}{\frac{\tilde{\mathbf{M}}_{1,1}}{\rho\left(s\right)}ds}=-k_{f}x_{0}\int^{\frac{\left(2m+1\right)\pi}{2\sqrt{k_{f}}}}_{0}{\frac{\cos^{2}\left(\sqrt{k_{f}}s\right)}{\left(1+k_{f}x^{2}_{0}\sin^{2}\left(\sqrt{k_{f}}s\right)\right)^{\frac{3}{2}}}ds} \\ \\
=\left(2m+1\right)\frac{K\left(-k_{f}x^{2}_{0}\right)-E\left(-k_{f}x^{2}_{0}\right)}{\sqrt{k_{f}}x_{0}},
\end{array}
\label{eq:spec7}
\end{equation}

\noindent where $K$ and $E$ are complete elliptic integrals of the first and second kind respectively. Eq. \ref{eq:spec7} only equates to zero when $x_{0}=0$; therefore $\sqrt{k_{f}}l_{q}=\left(2m+1\right)\pi/2$ leads to the trivial solution that $x_{0}=x'_{0}=0$.

For the case where $\sqrt{k_{f}}l_{q}=m\pi$, except for the trivial case where $\sqrt{k_{f}}l_{q}=0$, the quadrupole can be divided into three smaller quadrupoles such that

\begin{equation}
\begin{array}{l}
\mathbf{N}=\left(\begin{smallmatrix} \cos\left(2m\pi\right) & \frac{\sin\left(2m\pi\right)}{\sqrt{k_{f}}} \\
-\sqrt{k_{f}}\sin\left(2m\pi\right) & \cos\left(2m\pi\right) \end{smallmatrix}\right)=\left(\begin{smallmatrix} 0 & \frac{1}{\sqrt{k_{f}}} \\
-\sqrt{k_{f}} & 0 \end{smallmatrix}\right)\times \\ \left(\begin{smallmatrix} \cos\left(\left(2m-1\right)\pi\right) & \frac{\sin\left(\left(2m-1\right)\pi\right)}{\sqrt{k_{f}}} \\
-\sqrt{k_{f}}\sin\left(\left(2m-1\right)\pi\right) & \cos\left(\left(2m-1\right)\pi\right) \end{smallmatrix}\right)\left(\begin{smallmatrix} 0 & \frac{1}{\sqrt{k_{f}}} \\
-\sqrt{k_{f}} & 0 \end{smallmatrix}\right)
\end{array}.
\label{eq:spec8}
\end{equation}

We have proven that the only anti-symmetric solution for the central quadrupole is the trivial case where $x=x'=0$. Hence if we determine an initial trajectory which can produce this result after passing through the first quadrupole, we find

\begin{equation}
\begin{pmatrix}
0 & \frac{1}{\sqrt{k_{f}}} \\
-\sqrt{k_{f}} & 0 \end{pmatrix} \begin{pmatrix}
x_{0} \\
x'_{0} \end{pmatrix}=\begin{pmatrix}
\frac{x'_{0}}{\sqrt{k_{f}}} \\
-\sqrt{k_{f}}x_{0} \end{pmatrix}=\begin{pmatrix}
0 \\
0 \end{pmatrix}.
\label{eq:spec9}
\end{equation}

Therefore the only solution for the anti-symmetric focusing singlet is the trivial case where $x_{0}=x'_{0}=0$.

\subsection{$\mathbf{\tan\left(\sqrt{k_{f}}l_{q}\right)=\sqrt{k_{f}}L_{dr}/2}$ for the anti-symmetric focusing doublet}

From Eq. \ref{eq:bdoub3}, using the fact that $\det\left(\mathbf{N}\right)=1$, the transfer matrix for a doublet can be expressed as

\begin{equation}
\begin{array}{l}
\mathbf{N}= \\
\left(\begin{smallmatrix}\left(2\mathbf{M}_{1,1}+L_{dr}\mathbf{M}_{2,1}\right)\mathbf{M}_{1,1}-1 & \left(2\mathbf{M}_{1,1}+L_{dr}\mathbf{M}_{2,1}\right)\mathbf{M}_{1,2}+L_{dr} \\
\left(2\mathbf{M}_{1,1}+L_{dr}\mathbf{M}_{2,1}\right)\mathbf{M}_{2,1} & \left(2\mathbf{M}_{1,1}+L_{dr}\mathbf{M}_{2,1}\right)\mathbf{M}_{1,1}-1 \end{smallmatrix}\right)
\end{array}.
\label{eq:res4}
\end{equation}

From Eq. \ref{eq:bdoub15}, this can be simplified to

\begin{equation}
\mathbf{N}=\begin{pmatrix} -1 & L_{dr} \\
0 & -1 \end{pmatrix},
\label{eq:res5}
\end{equation}

\noindent hence we obtain the requirements for the anti-symmetric dispersion function

\begin{equation}
\begin{array}{l}
-D_{x,0}+L_{dr}D'_{x,0}+D_{q,doub}=-D_{x,0} \\
-D'_{x,0}+D'_{q,doub}=D'_{x,0}
\end{array},
\label{eq:res6}
\end{equation}

\noindent which implies that $-\frac{D_{q,doub}}{L_{dr}}=D'_{x,0}=\frac{D'_{q,doub}}{2}$ and therefore $D_{q,doub}\propto D'_{q,doub}$. But we have previously shown that this implies that $\det\left(\mathbf{N}\right)=0$, which is contradictory because we know that $\det\left(\mathbf{N}\right)=1$; therefore there are no non-trivial solutions for the anti-symmetric focusing doublet.

\section{\label{sec:nonlinear}Nonlinear Extension}

Having proven that no linear solution exists which can simultaneously close multiple orbit bumps and dispersion functions, we can consider the case where higher order multipoles are used to create a nonlinear optical system. For a particle bunch traveling on-axis through a multipole, a particle displaced by $x$ from the centroid of the bunch will experience a magnetic field given as

\begin{equation}
\mathbf{B}_{y}=\frac{pc}{e}k_{n}x^{n},
\label{eq:nl2}
\end{equation}

\noindent where $k_{0}$ represents the dipole term, $k_{1}$ the quadrupole term and so forth. $k_{n}$ can be defined as

\begin{equation}
k_{n}=\frac{e}{pc}\frac{\partial^{n}\mathbf{B}_{y}}{\partial{x}^{n}}.
\label{eq:nl1}
\end{equation}

If we now consider that the bunch centroid is off-axis by a distance $\delta{x}$ in the $x$-axis, then the magnetic field experienced by a particle displaced a distance $x$ from the bunch centroid will be

\begin{equation}
\mathbf{B}_{y}=\frac{p}{c}k_{n}\left(x+\delta{x}\right)^{n}=\frac{p}{c}k_{n}\sum^{n}_{k=0}{\begin{pmatrix}
n \\
k \end{pmatrix}x^{n-k}\delta{x}^{k}},
\label{eq:nl3}
\end{equation}

\noindent where $\begin{pmatrix}n \\k \end{pmatrix}=\frac{n!}{k!\left(n-k\right)!}$.

As we are only considering the linear transfer matrix, we only need to consider the dipole and quadrupole terms from Eq. \ref{eq:nl3}, which we shall define as $\tilde{K}$

\begin{equation}
\tilde{K}=nk_{n}\delta{x}^{n-1}+\frac{k_{n}\delta{x}^{n}}{x}.
\label{eq:nl4}
\end{equation}

To determine the equation of motion for a particle traveling through a multipole, we can exploit Hill's Equation

\begin{equation}
x''+K\left(s\right)x=0.
\label{eq:nl5}
\end{equation}

Substituting Eq. \ref{eq:nl4} into Eq. \ref{eq:nl5} gives the following equation of motion

\begin{equation}
x''+\left(nk_{n}\delta{x}^{n-1}\left(x,x',s\right)+\frac{k_{n}\delta{x}^{n}\left(x,x',s\right)}{x}\right)x=0.
\label{eq:nl6}
\end{equation}

The position of the bunch centroid, $\delta{x}$, varies with the longitudinal position, $s$, but also with the phase space coordinates, $\left(x,x'\right)$; therefore Hill's Equation becomes a nonlinear differential equation with no analytical solution. To overcome this, we will consider the multipole as a series of slices; such that $\tilde{K}$ varies a negligible amount in each slice. It should be noted though that we do not use the thin lens approximation, but rather, we assume that $\delta{\tilde{K}}\approx{0}$ through the slice.

If we consider $\tilde{K}$ for a quadrupole in Eq. \ref{eq:nl4}, we obtain

\begin{equation}
\tilde{K}_{1}x=k_{1}x+k_{1}\delta{x}.
\label{eq:nl7}
\end{equation}

However, for a higher order multipole, we can rearrange Eq. \ref{eq:nl4} as

\begin{equation}
\begin{array}{c}
\tilde{K}_{n}x=\left(nk_{n}\delta{x}^{n-1}\right)x+\left(nk_{n}\delta{x}^{n-1}\right)\frac{\delta{x}}{n} \\
=\tilde{k}_{1}x+\tilde{k}_{1}\frac{\delta{x}}{n}
\end{array},
\label{eq:nl8}
\end{equation}

\noindent where $\tilde{k}_{1}$ is the effective quadrupole term for the trajectory through the multipole. Therefore traveling off-axis through a multipole of order $n$ with a transverse displacement $\delta{x}$ is equivalent to traveling off-axis through a quadrupole with a transverse displacement $\frac{\delta{x}}{n}$. As previously shown for the linear case, the trivial solution where $x=x'=0$ is the only possible solution. For the nonlinear case, the result from Eq. \ref{eq:nl8} implies that as the order of the multipole increases, one can design an optical system which asymptotically converges to a closed solution for the orbit bump and dispersion; although no perfect solution exists for any system of linear or nonlinear optics.

From Eq. \ref{eq:nl4}, the fractional error on $\tilde{K}$ in terms of beam jitter, $\sigma_{x}$ can be calculated as

\begin{equation}
\begin{array}{c}
\frac{\sigma_{\tilde{K}}}{\tilde{K}}=\frac{\partial{\tilde{K}}}{\partial{\delta{x}}}\frac{\sigma_{x}}{\tilde{K}}=n\frac{\sigma_{\delta{x}}}{\delta{x}}\left(1-\frac{x}{nx+\delta{x}}\right)\\
\approx{n}\frac{\sigma_{\delta{x}}}{\delta{x}}
\end{array}.
\label{eq:nl9}
\end{equation}

Therefore the tolerance on fractional beam jitter scales as $1/n$ for a multipole of order $n$. Hence for higher order multipoles, the jitter tolerances become increasingly stringent. This increasingly tight tolerance on the beam jitter would likely limit the maximum order of a multipole which can be used for such a system before the residual dispersion introduced by the beam jitter exceeds the residual dispersion of an off-axis orbit bump.

\section{Summary}

In this paper we have shown that there is no possible linear solution to simultaneously close orbit and dispersion functions. We showed that if a solution exists then it must be possible to create a symmetric lattice which is also a solution. For a symmetric lattice, both the orbit and dispersion must be either symmetric or anti-symmetric about the midpoint of the lattice. This allows us to investigate just the central region of the lattice to determine whether a solution is possible. By considering a quadrupole singlet at the centre of the injection region, we are able to draw conclusions about any lattice consisting of an odd number of quadrupoles. Similarly by considering a doublet at the centre, we are able to draw conclusions about any lattice consisting of an even number of quadrupoles. By considering a quadrupole singlet as the special case of a quadrupole doublet with a drift length $L_{dr}=0$, we are able to investigate all cases and show that no non-trivial linear solutions exist.

After proving that no linear solution exists, we were able to extend the proof to nonlinear optical systems. By considering the multipole terms experienced by an off-axis beam and linearising Hill's Equation, we were able to determine the effective linear equation of motion of an off-axis particle. By relating this to the proof for linear optics, we were able to show that no non-trivial nonlinear optics exist; thus completing the proof that no solution exists to simultaneously correct multiple local orbit bumps and dispersion functions with linear or nonlinear optics.

\begin{acknowledgments}
The authors wish to thank Dr Oznur Mete for her useful insight and suggestions with regards to this paper.
\end{acknowledgments}

\bibliography{off_axis_prstab}

\providecommand{\noopsort}[1]{}\providecommand{\singleletter}[1]{#1}%
\begin{thebibliography}{2}%
\makeatletter
\providecommand \@ifxundefined [1]{%
 \@ifx{#1\undefined}
}%
\providecommand \@ifnum [1]{%
 \ifnum #1\expandafter \@firstoftwo
 \else \expandafter \@secondoftwo
 \fi
}%
\providecommand \@ifx [1]{%
 \ifx #1\expandafter \@firstoftwo
 \else \expandafter \@secondoftwo
 \fi
}%
\providecommand \natexlab [1]{#1}%
\providecommand \enquote  [1]{``#1''}%
\providecommand \bibnamefont  [1]{#1}%
\providecommand \bibfnamefont [1]{#1}%
\providecommand \citenamefont [1]{#1}%
\providecommand \href@noop [0]{\@secondoftwo}%
\providecommand \href [0]{\begingroup \@sanitize@url \@href}%
\providecommand \@href[1]{\@@startlink{#1}\@@href}%
\providecommand \@@href[1]{\endgroup#1\@@endlink}%
\providecommand \@sanitize@url [0]{\catcode `\\12\catcode `\$12\catcode
  `\&12\catcode `\#12\catcode `\^12\catcode `\_12\catcode `\%12\relax}%
\providecommand \@@startlink[1]{}%
\providecommand \@@endlink[0]{}%
\providecommand \url  [0]{\begingroup\@sanitize@url \@url }%
\providecommand \@url [1]{\endgroup\@href {#1}{\urlprefix }}%
\providecommand \urlprefix  [0]{URL }%
\providecommand \Eprint [0]{\href }%
\providecommand \doibase [0]{http://dx.doi.org/}%
\providecommand \selectlanguage [0]{\@gobble}%
\providecommand \bibinfo  [0]{\@secondoftwo}%
\providecommand \bibfield  [0]{\@secondoftwo}%
\providecommand \translation [1]{[#1]}%
\providecommand \BibitemOpen [0]{}%
\providecommand \bibitemStop [0]{}%
\providecommand \bibitemNoStop [0]{.\EOS\space}%
\providecommand \EOS [0]{\spacefactor3000\relax}%
\providecommand \BibitemShut  [1]{\csname bibitem#1\endcsname}%
\let\auto@bib@innerbib\@empty
\bibitem [{\citenamefont {CLIC}(2012)}]{CLICCDR}%
  \BibitemOpen
  \bibfield  {author} {\bibinfo {author} {\bibnamefont {CLIC}},\ }\href@noop {}
  {\emph {\bibinfo {title} {A Multi-TeV Linear Collider based on CLIC
  Technology}}}\ (\bibinfo  {publisher} {CERN-12-007},\ \bibinfo {year}
  {2012})\BibitemShut {NoStop}%
\bibitem [{\citenamefont {Apsimon}\ and\ \citenamefont
  {Esberg}(2014)}]{CLICnote}%
  \BibitemOpen
  \bibfield  {author} {\bibinfo {author} {\bibfnamefont {R.}~\bibnamefont
  {Apsimon}}\ and\ \bibinfo {author} {\bibfnamefont {J.}~\bibnamefont
  {Esberg}},\ }\href@noop {} {\emph {\bibinfo {title} {Proof of the
  Nonexistence of a Linear Solution for the CR2 Injection Region of the CLIC
  Drive Beam}}}\ (\bibinfo  {publisher} {CLIC-Note-1035},\ \bibinfo {year}
  {2014})\BibitemShut {NoStop}%
\end{thebibliography}%

\end{document}